\documentclass[12pt,preprint]{aastex}
\usepackage{epsfig}

\shorttitle{Swift GRB Mission}

\begin{document}

\title{The Swift Gamma-Ray Burst Mission}

\author{N.~Gehrels\altaffilmark{1}, G~ Chincarini\altaffilmark{2,3},
P.~Giommi\altaffilmark{4}, K.~O.~Mason\altaffilmark{5},
J.~A.~Nousek\altaffilmark{6}, A.~A.~Wells\altaffilmark{7},
N.~E.~White\altaffilmark{1}, S.~D.~Barthelmy\altaffilmark{1},
D.~N.~Burrows\altaffilmark{6}, L.~R.~Cominsky\altaffilmark{8},
K.~C.~Hurley\altaffilmark{9}, F.~E.~Marshall\altaffilmark{1},
P.~M\'{e}sz\'{a}ros\altaffilmark{6}, P.~W.~A.~Roming\altaffilmark{6},
L.~Angelini\altaffilmark{1,10}, L.~M.~Barbier\altaffilmark{1},
T.~Belloni\altaffilmark{2}, P.~T.~Boyd\altaffilmark{1,11}, 
S.~Campana\altaffilmark{2},P.~A.~Caraveo\altaffilmark{12}, 
M.~M.~Chester\altaffilmark{6},O.~Citterio\altaffilmark{2}, 
T.~L.~Cline\altaffilmark{1}, M.~S.~Cropper\altaffilmark{5}, 
J.~ R.~Cummings\altaffilmark{1,13}, A.~J.~Dean\altaffilmark{14}, 
E.~D.~Feigelson\altaffilmark{6}, E.~E.~Fenimore\altaffilmark{15}, 
D.~A.~Frail\altaffilmark{16}, A.~S.~Fruchter\altaffilmark{17}, 
G.~P.~Garmire\altaffilmark{6}, K.~ Gendreau\altaffilmark{1}, 
G.~Ghisellini\altaffilmark{2}, J.~Greiner\altaffilmark{18}, 
J.~E.~Hill\altaffilmark{6}, S.~D.~Hunsberger\altaffilmark{6}, 
H.~A.~Krimm\altaffilmark{1,10}, S.~R.~Kulkarni\altaffilmark{19}, 
P.~Kumar\altaffilmark{20}, F.~Lebrun\altaffilmark{21}, 
N.~M.~Lloyd-Ronning\altaffilmark{22}, C.~B.~Markwardt\altaffilmark{1,23}, 
B.~J.~Mattson\altaffilmark{1,23,24}, R.~F.~Mushotzky\altaffilmark{1}, 
J.~P.~Norris\altaffilmark{1}, B.~Paczynski\altaffilmark{25}, 
D.~M.~Palmer\altaffilmark{15}, H.-S.~Park\altaffilmark{26}, 
A.~M.~Parsons\altaffilmark{1}, J.~Paul\altaffilmark{21}, 
M.~J.~Rees\altaffilmark{27}, C.~S.~Reynolds\altaffilmark{23}, 
J.~E.~Rhoads\altaffilmark{17}, T.~P.~Sasseen\altaffilmark{28}, 
B.~E.~Schaefer\altaffilmark{20}, A.~T.~Short\altaffilmark{29}, 
A.~P.~Smale\altaffilmark{1,10}, I.~A.~Smith\altaffilmark{30}, 
L.~Stella\altaffilmark{31}, M.~Still\altaffilmark{1,10}, 
G.~Tagliaferri\altaffilmark{2}, T.~Takahashi\altaffilmark{32,33},
M.~Tashiro\altaffilmark{32,34}, L.~K.~Townsley\altaffilmark{6},
J.~Tueller\altaffilmark{1}, M.~J.~L.~Turner\altaffilmark{29},
M.~Vietri\altaffilmark{35}, W.~Voges\altaffilmark{18},
M.~J.~Ward\altaffilmark{29}, R.~Willingale\altaffilmark{7},
F.~M.~Zerbi\altaffilmark{2}, W.~W.~Zhang\altaffilmark{1}}

\altaffiltext{1}{NASA/Goddard Space Flight Center Greenbelt, MD}
\altaffiltext{2}{Osservatorio Astronomico di Brera, Milano, Italy}
\altaffiltext{3}{Universita degli Studi di Milano Bicocca}
\altaffiltext{4}{ASI Science Data Center, ASI, Roma, Italy}
\altaffiltext{5}{Mullard Space Science Laboratory, University College London,
	Dorking, UK}
\altaffiltext{6}{Department of Astronomy and Astrophysics, Pennsylvania State
	University, University Park, PA}
\altaffiltext{7}{Space Research Centre, University of Leicester, Leicester, UK}
\altaffiltext{8}{Department of Physics and Astronomy, Sonoma State University,
	Rohnert Park, CA}
\altaffiltext{9}{University of California Space Sciences Laboratory, Berkeley,
	CA}
\altaffiltext{10}{Universities Space Research Association, Columbia, MD}
\altaffiltext{11}{Joint Center for Astrophysics, University of Maryland, Baltimore County, MD}
\altaffiltext{12}{Istituto di Astrofisica 
Spaziale e Fisica Cosmica,  CNR, Milano,
	Italy}
\altaffiltext{13}{National Research Council, Washington, DC}
\altaffiltext{14}{Department of Physics and 
Astronomy, University of Southampton
	Highfield, Southampton, UK}
\altaffiltext{15}{Los Alamos National Laboratory, Los Alamos, NM}
\altaffiltext{16}{National Radio Astronomy Observatory, Socorro, NM}
\altaffiltext{17}{Space Telescope Science Institute, Baltimore, MD}
\altaffiltext{18}{Max Planck Institut fÄr Extraterrestrische Physik, Garching,
	Germany}
\altaffiltext{19}{Department of Astronomy, California Institute of Technology,
	Pasadena, CA}
\altaffiltext{20}{Department of Astronomy, University of Texas at Austin,
	Austin, TX}
\altaffiltext{21}{CEA, DSM/DAPNIA/SAP, Centre 
d'Etudes de Saclay, Cedex, France}
\altaffiltext{22}{Canadian Institute for Theoretical Astrophysics McClennan
Labs, University of Toronto, Toronto, Ontario, Canada}
\altaffiltext{23}{Department of Astronomy, 
University of Maryland, College Park,
	MD}
\altaffiltext{24}{L-3 Communications EER, Chantilly, VA}
\altaffiltext{25}{Princeton University Observatory, Princeton, NJ}
\altaffiltext{26}{Lawrence Livermore National Laboratory, Livermore, CA}
\altaffiltext{27}{Institute of Astronomy, University of Cambridge, Cambridge,
	England, UK}
\altaffiltext{28}{Department of Physics, University of California, Santa
	Barbara, CA}
\altaffiltext{29}{Physics and Astronomy Department, University of Leicester,
	Leicester, UK}
\altaffiltext{30}{Department of Physics and 
Astronomy, Rice University, Houston,
	TX}
\altaffiltext{31}{Osservatorio Astronomico di Roma, Monteporzio Catone, Italy}
\altaffiltext{32}{Institute of Space and 
Astronautical Science, Kanagawa, Japan}
\altaffiltext{33}{Department of Physics, University of Tokyo, Tokyo, Japan}
\altaffiltext{34}{Department of Physics, Saitama 
University, Sakura, Saitama, Japan}
\altaffiltext{35}{Arcetri Astrophysical Observatory, Firenze, Italy}

\begin{abstract}
The Swift mission, scheduled for launch in early 2004, is a
multiwavelength observatory for gamma-ray burst (GRB) astronomy.  It is
the first-of-its-kind autonomous rapid-slewing satellite for transient
astronomy and pioneers the way for future rapid-reaction and
multiwavelength missions.  It will be far more powerful than any
previous GRB mission, observing more than 100 bursts per year and
performing detailed X-ray and UV/optical afterglow observations spanning
timescales from 1 minute to several days after the burst.  The
objectives are to: 1) determine the origin of GRBs; 2) classify GRBs and
search for new types; 3) study the interaction of the ultra-relativistic
outflows of GRBs with their surrounding medium; and 4) use GRBs to study
the early universe out to $z > 10$.  The mission is being developed by a
NASA-led international collaboration.  It will carry three instruments:
a new-generation wide-field gamma-ray (15-150 keV) detector that will
detect bursts, calculate 1-4 arcmin positions, and trigger autonomous
spacecraft slews; a narrow-field X-ray telescope that will give 5 arcsec
positions and perform spectroscopy in the 0.2 to 10 keV band; and a
narrow-field UV/optical telescope that will operate in the 170-600 nm
band and provide 0.3 arcsec positions and optical finding charts.
Redshift determinations will be made for most bursts.  In addition to
the primary GRB science, the mission will perform a hard X-ray survey to
a sensitivity of $\sim 1$ mCrab ($\sim 2 \times 10^{-11}$ erg cm$^{-2}$
s$^{-1}$ in the 15-150 keV band), more than an order of magnitude better
than HEAO A-4.  A flexible data and operations system will allow rapid
follow-up observations of all types of high-energy transients, with
rapid data downlink and uplink available through the NASA TDRSS system.
Swift transient data will be rapidly distributed to the astronomical
community and all interested observers are encouraged to participate in
follow-up measurements.  A Guest Investigator program for the mission
will provide funding for community involvement.  Innovations from the
Swift program applicable to the future include: 1) a large-area
gamma-ray detector using the new CdZnTe detectors; 2) an autonomous
rapid slewing spacecraft; 3) a multiwavelength payload combining
optical, X-ray, and gamma-ray instruments; 4) an observing program
coordinated with other ground-based and space-based observatories; and
5) immediate multiwavelength data flow to the community.  The mission is
currently funded for 2 years of operations and the spacecraft will have
a lifetime to orbital decay of $\sim 8$ years.
\end{abstract}


\section{Introduction}
\label{section:introduction}

\begin{figure}[ht]
   \figurenum{1}
   \epsscale{0.3}
   \plotone{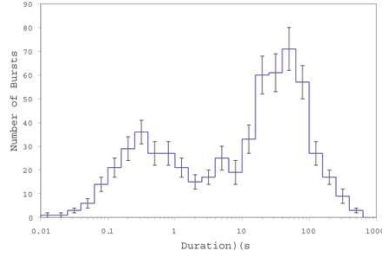}
   \caption{Burst duration versus number for GRBs detected by BATSE.  The
   two peaks occur at $\sim 0.3$ second and $\sim 30$ seconds (based on
   \citet{meegan1996}).}
   \label{fig1}
\end{figure}

Gamma-ray bursts (GRBs) were discovered in the late 1960s in data from
the Vela satellites \citep{klebesadel1973}.  Tremendous progress has
been made in their understanding over the past thirty years and
particularly since 1997.  We know that they are bright ($\sim$few
photons cm$^{-2}$ s$^{-1}$ flux in the 50-300 keV band) flashes of gamma
rays that are observable at Earth approximately once per day.  The BATSE
instrument on the Compton Gamma-Ray Observatory showed that they are
distributed isotropically over the sky \citep{briggs1996} and show a
deficit of very faint bursts \citep{paciesas1999}.  GRBs have durations
ranging from milliseconds to tens of minutes, with a bimodal
distribution showing clustering at $\sim 0.3$ second (short bursts) and
$\sim 30$ seconds (long bursts) as shown in Figure~\ref{fig1}. For long
bursts, the discovery by BeppoSAX \citep{costa1997} and ground-based
observers \citep{vanparadijs1997, frail1997} of X-ray through radio
afterglow allowed redshifts to be measured and host galaxies to be
found, proving a cosmological origin.  In contrast, little is known
about the short bursts and their afterglows \citep{hurley2002}.  With
typical redshifts of $z \sim 1$, the gamma-ray flash corresponds to a
huge instantaneous energy release of $10^{51}-10^{52}$ ergs (assuming
the radiation is beamed into $\sim 0.1$ steradian).  GRBs are probably
related to black hole formation, possibly related to endpoints of
stellar evolution, and definitely bright beacons from the high redshift
universe (see \citet{vanparadijs2000} for a review).  The recent
afterglow discoveries have illustrated that multiwavelength studies are
the key to our further understanding of GRBs.  Swift is designed
specifically to study GRBs and their afterglow in multiple wavebands.
It will perform sensitive X-ray and optical afterglow observations of
hundreds of GRBs on all timescales, from a minute after the burst
detection to hours and days later.  Since afterglows fade quickly,
typically as $t^{-1}$ or $t^{-2}$, Swift's rapid $\sim 1$ minute
response will allow observations when the emissions are orders of
magnitude brighter than the current few-hour response capabilities.

\begin{figure}[ht]
   \figurenum{2}
   \epsscale{0.3}
   \plotone{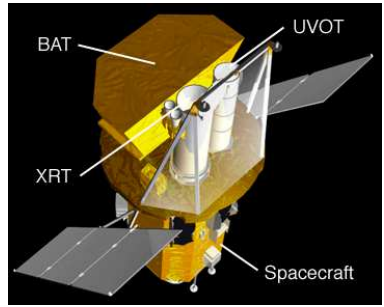}
   \caption{The Swift satellite.}
   \label{fig2}
\end{figure}

Swift is a medium-sized explorer (MIDEX) mission selected by NASA for
launch in early 2004.  The hardware is being developed by an
international team from the USA, the United Kingdom, and Italy, with
additional scientific involvement in France, Japan, Germany, Denmark,
Spain and South Africa.  The primary scientific objectives are to
determine the origin of GRBs and to pioneer their usage as probes of the
early universe.  Swift's Burst Alert Telescope (BAT), will search the
sky for new GRBs and, upon discovery, will trigger an autonomous
spacecraft slew to bring the burst into the X-Ray Telescope (XRT) and
Ultraviolet/Optical Telescope (UVOT) fields of view (FOVs).  Such
autonomy will allow Swift to perform X-ray and UV/optical observations
of $> 100$ bursts per year within 20-70 seconds of a burst detection.
Figure~\ref{fig2} shows a drawing of the Swift spacecraft, and
Table~\ref{tab1} summarizes Swift's mission characteristics.
Tables~\ref{tab2}, \ref{tab3}, and~\ref{tab4} list the parameters of the
three instruments.

\section{Key Science}
\label{section:science}

The Swift mission provides the capability to answer four key GRB science
questions:  What are the progenitors of GRBs?  Are there different
classes of bursts with unique physical processes at work?  How does the
blastwave evolve and interact with its surroundings?  What can GRBs tell
us about the early Universe?  In addition, the mission will carry out a
broad program of non-GRB science.

\subsection{GRB Progenitors}
\label{section:science:progenitors}

Three parameters are necessary for the determination of GRB progenitors:
the total energy released, the nature of the host galaxy (if one
exists), and the location of the burst within the host galaxy.  The
angular resolution of the XRT and UVOT allows precise location of the
BAT-discovered bursts, yielding measurements of these parameters for
hundreds of bursts.

To measure the total energetics of a burst, reliable redshifts are
needed.  Ideally, this should be done independently for the afterglow
and proposed host galaxy to rule out a chance juxtaposition
\citep{hogg1999}. Swift's UV grisms and filters can make redshift
determinations by searching for the Ly-$\alpha$ cutoff in the UV and
eliminate the $1.3 < z < 2.5$ deadband of current observations during
the early phase of the afterglow.  For GRBs with afterglow optical
brightness of $m < 17$, the UVOT grism will perform spectroscopy between
170 and 600 nm with $\lambda / \Delta \lambda \; \sim 200$ resolution.
In addition, illumination of the immediate (100 pc) environment by the
initial burst is expected to cause time varying optical, UV and X-ray
lines and edges within the first hour \citep{perna1998, meszaros1998},
evolving at later times to give abundance information on the
circumstellar medium \citep{reeves2002, butler2003}.  Swift's rapid
response will allow a search for the expected X-ray lines and, thereby
also provide a direct redshift measure from the X-ray afterglow.

The UVOT will provide positions of $< 0.3$ arcsec accuracy by using
background stars to register the field.  This position will give unique
host galaxy identifications and allow later comparison with HST fields
to determine the burst's position within the galaxy.

In some events, there may be no optical afterglow visible due to dust
extinction surrounding the GRB site \citep{lamb2003} or Lyman cutoff of
a high redshift event.  This is not likely to occur often with UVOT
observing to sensitivities of 24$^{th}$ magnitude immediately following
the burst, as described below, but when it does occur it will indicate a
high priority GRB.  In such cases, the XRT 5.0 arcsec positions will be
crucial, allowing unique identification of the candidate galaxy down to
$m_{R} \sim 26$ and rapid ground-based IR follow-up.  Observations with
Chandra made within a couple of days for a selection of these events
will give sub-arcsec positions within the Swift 5.0 arcsec error circle.

\subsection{Blastwave Interaction}
\label{section:science:blastwave}

The GRB afterglow is thought to be produced by the interaction of an
ultra-relativistic blastwave with the interstellar medium (ISM) or
intergalactic medium (IGM).  The blastwave model \citep{rees1992}
predicts a series of stages as the wave slows.  A key prediction is a
break in the spectrum that moves from the gamma to optical band, and is
responsible for the power law decay of the source flux
\citep{meszaros1997}.  This break moves through the X-ray band in a few
seconds, but takes up to 1000 s to reach the optical; thus, observations
within the first 1000 s in the optical and UV are critical.  While it
now seems likely that all the long GRBs have X-ray afterglows, not all
have bright optical or radio afterglow (at least after several hours).
While this may be due to optical extinction, it is possible that in some
cases the optical (and X-ray) afterglow is present but decays much more
rapidly \citep{groot1998, pandey2003}, perhaps as a function of the
density of the local environment \citep{piran1998}.  Prompt high-quality
X-ray, UV, and optical observations over the first minutes to hours of
the afterglow are crucial to resolving this question.  Continuous
monitoring is important since model-constraining flares can occur in the
decaying emission.

Swift's capability to detect X-ray spectral lines and edges will provide
a wealth of information about the afterglow mechanism and sites,
including density, ionization, elemental abundance, and outflow
characteristics.  Swift will enable the observation of lines during the
early bright phase of the afterglow, during which they are best
detected.

Star forming regions are embedded in large columns of neutral gas and
dust.  The presence of extinction can be readily determined by multiband
photometry in the optical and IR.  The simultaneous detection of high
X-ray absorption, coupled with photometric E (\bv) measurements with
Swift, will determine whether dust and gas are present.  Continuous
monitoring over the first few hours to days will indicate whether dust
is building up (due to condensation out of an expanding hot wind) or
disappearing (due to ablation and evaporation).

\subsection{Classes of GRB}
\label{section:science:classes}

By determining the redshift, location, and afterglow properties of many
hundreds of bursts, Swift will determine whether or not sub-classes of
GRBs exist and what physical differences cause the classes.  The current
evidence for sub-classes is a bimodal distribution of burst duration, a
possible correlation between hardness and $\log \: N$-$\log \: P$ shape,
the consistency of the $V / V_{max}$ of some short bursts with a
Euclidean distribution, the detection of X-ray rich events, the
non-detection of optical emission from ``dark'' bursts and a possible
separate population of long-lag, low-luminosity GRBs.  However, it is
not clear whether these differences are real or artifacts of the
distribution function of GRB properties such as beaming angle, density
of the local medium or initial energy injection.  The main reason for
current confusion is that no standard candle exists for GRBs, although
recent work shows that when collimation angles are taken into account
the total energy seems to be more narrowly distributed than the fluence
\citep{panaitescu2001, frail2001}.  Swift will remedy this confusion by
directly measuring distance through redshift, thereby giving an exact
determination of the GRB luminosity function.

Since BeppoSAX was not able to accurately locate bursts shorter than
$\sim 1$ second and because short bursts tend to have hard spectra to
which HETE-2 is not sensitive, we have little data on the nature of
afterglow for the short class of GRBs (see Figure~\ref{fig1}).  Swift
will be sensitive to the shortest events, so will provide better
coverage of these events than has been possible with current missions.

Should Swift discover GRBs with no X-ray or UV/optical afterglow, the
BAT will still provide positions of 1-4 arcmin, which is sufficient to
look for radio or IR counterparts.  Only the rapid response of Swift
will be able to identify such a new and elusive subclass of GRB event.

There is growing evidence of an association of GRBs with supernova
explosions \citep{bloom1999, woosley1999, galama1998, germany2000,
reichart1999, dado2002, stanek2003, hjorth2003, dellavalle2003}.  For
such associations, the UVOT will provide unique and unprecedented
coverage of the optical and UV light curve during the early stage.

\subsection{GRBs as Astrophysical Tools}
\label{section:science:astrophycial_tools}

Since the lifetime of massive star progenitors is short compared to the
age of the universe at $z < 5$, the cosmic GRB rate should be
approximately proportional to the star formation rate.  The cosmic rate
of massive star formation is at present controversial.  Estimates that
star formation peaks at $z \sim $1-2 and declines sharply at high
redshifts have been reported \citep{madau1996}.  However, IR
\citep{blain1999, rowanrobinson1997} and X-ray cluster
\citep{mushotzky1997} results show a considerably higher rate in dust
enshrouded galaxies at higher redshifts.  Swift, by obtaining a large
sample of GRBs over a wide range of fluences and redshifts, will provide
valuable information on the evolution of star formation in the Universe
\citep{lamb2000, bromm2002, lloydronning2002}. Also, because the X-ray
flux does not depend greatly on the line of sight column, these results
will be independent of absorption.  The star formation rates in the
Swift-selected host galaxies can independently be estimated, for example
using sub-millimeter and radio observations \citep{barnard2003,
berger2003}.

GRBs are the most luminous objects we know of in the Universe, and, as
such, provide a unique opportunity to probe the IGM and ISM of the host
galaxies via measurement of absorption along the line of sight
\citep{lamb2000, fiore2000}.  Depending on evolution, GRBs might
originate from redshifts up to $\sim 15$ and have a median redshift $>
2$, larger than that of any other observable population.  By rapidly
providing both accurate positions and optical brightness, Swift will
enable the immediate follow-up of those GRBs bright enough for high
resolution optical absorption line spectroscopy at redshifts high enough
to study the reionization of the IGM \citep{miraldaescude1998}.  This
information on the high-z Ly-$\alpha$ forest will be unique because
there are currently no known bright ($m < 17$) galaxies or quasars at $z
>  6.5$ \citep{fan2001, lamb2000}.

\subsection{Non-GRB Science}
\label{section:science:nongrb}

\subsubsection{Hard X-ray Survey}
\label{section:science:hardsurvey}

The BAT will produce the most sensitive hard X-ray survey ever made.
Since no all-sky survey is planned by INTEGRAL, Swift's survey will be
unmatched.  Assuming uniform coverage to estimate sensitivities, the BAT
instrument will provide an exposure of $1.3 \times 10^{10}$ cm$^{2}$ s
for each sky pixel yielding a $5 \sigma$ statistical sensitivity of 200
mCrab in the 15-150 keV band.  In this energy 
range, our experiencewith coded mask instruments 
suggests that such deep exposures
will be systematics limited.  Although 2D coded apertures generally have
better systematics than the alternatives (modulation collimators,
Fourier grids, etc.), we estimate BAT detection sensitivity to be
systematics limited at $\sim 1$ mCrab at high galactic latitude ($> 45$
\degr) and $\sim 3$ mCrab when strong galactic sources are in the FOV.
These correspond to limits in the 15-150 keV band of $2 \times 10^{-11}$
erg cm$^{-2}$ s$^{-1}$ for high latitudes and $6 \times 10^{-11}$ erg
cm$^{-2}$ s$^{-1}$ for low latitudes.  Such levels correspond to $15
\sigma$ to $50 \sigma$ in a statistical sense. Swift's survey will be 17
times more sensitive for one third of the sky and 5 times more sensitive
for the rest than the best complete hard X-ray survey to date by HEAO
A-4, complete to 17 mCrab \citep{levine1984}.

\subsubsection{Active Galactic Nuclei}
\label{section:science:agn}

Recent studies with ASCA, Ginga, and BeppoSAX have shown the existence
of a large population of highly absorbed Seyfert 2 galaxies with
line-of-sight column densities $>10^{23}$ cm$^{-2}$.  The large column
makes the nuclei of these objects essentially invisible at optical and
soft X-ray wavelengths.  Detailed models \citep{madau1994, hasinger1997}
show such a population of highly absorbed AGN is needed to produce the
observed 30 keV bump in the hard X-ray background and that it comprises
about half of all AGN.  The only known method of detecting such objects
is an unbiased sky survey in the E $> 10$ keV band of sufficient
sensitivity to detect a large population.

The number of AGN observed at $> 1$ mCrab in the 2-10 keV range is $\sim
100$ \citep{piccinotti1982}, but AGN spectra are harder than that of the
Crab, so this corresponds to 200-300 sources at $> 1$ mCrab in the
10-100 keV band.  The factor for highly absorbed AGN raises this total
to 400-600 sources.  These estimates are consistent with scaling from
the 6 AGN in the HEAO A-4 survey.  Using 400 sources, we expect Swift to
detect $> 300$ AGN brighter than 1 mCrab within 45\degr of the Galactic
poles and $> 70$ AGN brighter than 3 mCrab in the rest of the sky.  More
than half will not have been identified in the ROSAT survey.  At this
time, only $\sim 20$ AGN have quality detections at energies $> 30$ keV
\citep{dermer1995, macomb1999}.

\subsubsection{Soft Gamma Repeaters}
\label{section:science:sgr}

The BAT will be sensitive to soft gamma repeaters (SGRs) because of its
capable short-burst trigger.  While the SGR bursts are shorter than the
Swift slew time, XRT and UVOT will perform sensitive searches
immediately after an SGR burst for X-ray and optical counterparts and
will likely be on-target when subsequent bursts occur.  Swift will not
only study existing SGRs, but will be able to discover new ones to help
complete the galactic SGR census.

\subsubsection{Rapid Reaction Science}
\label{section:science:rapidreaction}

The rapid reaction capability of the Swift observatory, using the TDRSS
uplink, will provide the unique ability to respond rapidly (within $\sim
1$ hour) with sensitive gamma-ray, X-ray, UV, and optical observations
to most events on the sky.  This includes targets of opportunity (ToOs)
for AGN flares, X-ray transients, pulsar glitches, outbursts from dwarf
novae, and stellar flares.  Highly variable black-hole binaries such as
GX~339-4 will be excellent targets for simultaneous high time resolution
multiwavelength observations for understanding the complex
accretion-outflow physics \citep{smith1999}.  The BAT is $> 10$ times
more sensitive as a monitor than BATSE and will initiate many of the
targets of opportunity.  Triggers from external sources will also be
possible.  There has never been a facility that can provide such rapid
multiwavelength follow-up to unpredictable events. As with any new
observation capability, the potential for serendipitous science return
is high.

\section{Swift Mission}
\label{section:mission}

The Swift mission was selected for Phase A study in January 1999 and
selected for flight in October 1999.  Swift's science payload consists
of three instruments mounted onto an optical bench.  These instruments,
the Burst Alert Telescope (BAT), X-Ray Telescope (XRT) and
UltraViolet/Optical Telescope (UVOT), are shown on the spacecraft in
Figure~\ref{fig2}, and their characteristics are listed in
Tables~\ref{tab2}-\ref{tab4}.

The spacecraft is provided by Spectrum Astro, based on their
flight-proven SA-200 bus.  The launch will be on a Delta 7320-10 in
early 2004 to a 22\degr-inclination, 600-km altitude orbit.  Swift has a
nominal lifetime of 2 years with a goal of 5 years and an orbital
lifetime of $\sim 8$ years.

Normal data will be downlinked in several passes each day over the
Italian Space Agency (ASI) ground station at Malindi, Kenya.  TDRSS will
be used to send burst alert messages to the ground.  Similarly,
information about bursts observed by other spacecraft will be uplinked
through TDRSS for evaluation by Swift's on-board figure-of-merit (FoM)
software (see Section~\ref{section:bat:technical}).

\begin{figure}[ht]
   \figurenum{3}
   \epsscale{0.3}
   \plotone{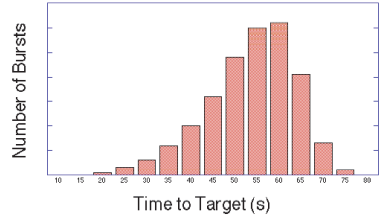}
   \caption{Simulated distribution of reaction time.  The time to target
   is 10 s plus the slew time.}
   \label{fig3}
\end{figure}

There will be no on-board propulsion system.  Pointing will be provided
by momentum wheels with momentum unloaded by magnetic torquers. Pointing
knowledge will be through the gyroscopes, star trackers, sun sensors,
and magnetometers.  The simulated distribution of Swift reaction times
is shown in Figure~\ref{fig3}.

The observation strategy for the early phases of the mission will be for
the XRT and UVOT to be almost constantly observing positions of
bursts previously detected by the BAT.  The most recently detected burst
will have priority (although this can be adjusted as science dictates). 
Typically 2 to 4 sources will be observed each orbit.  When a burst is
detected, an automated series of XRT and UVOT observations will be
performed lasting 20,000 s in exposure.  Following that time, additional
observations will be scheduled via ground planning.  For at least the
first months of the mission, all GRBs detected/imaged by BAT and within
allowed pointing constraints of the spacecraft will be slewed to.  Also,
priority will be given to GRB observations in this period with no time
specifically spent on secondary science; although, serendipitous science
such as the BAT sky survey and transient monitoring will occur
automatically as BAT awaits the next GRB trigger.

\section{Burst Alert Telescope} \label{section:bat}

The Burst Alert Telescope (BAT) is a highly sensitive, large FOV
instrument designed to provide critical GRB triggers and 4-arcmin
positions.  It is a coded-mask instrument with a 1.4 steradian
field-of-view (half coded).  The energy range is 15-150 keV for imaging
with a non-coded response up to 500 keV.  Within the first $\sim 10$
seconds of detecting a burst, the BAT will calculate an initial
position, decide whether the burst merits a spacecraft slew and, if
worthy, send the position to the spacecraft.

\begin{figure}[ht]
   \figurenum{4}
   \epsscale{0.3}
   \plotone{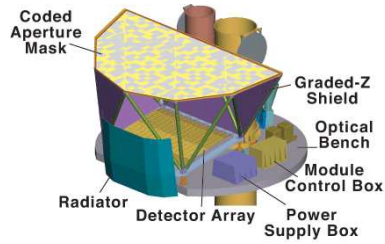}
   \caption{The Burst Alert Telescope cut away drawing showing the
   D-shaped coded mask, the CZT array, and the Graded-Z shielding.  The
   mask pattern is not to scale.}
   \label{fig4}
\end{figure}

In order to study bursts with a variety of intensities, durations, and
temporal structures, the BAT must have a large dynamical range and
trigger capabilities.  The BAT uses a two-dimensional coded mask and a
large area solid state detector array to detect weak bursts and has a
large FOV to detect a good fraction of bright bursts.  Since the BAT
coded FOV always includes the XRT and UVOT fields-of-view, long duration
gamma-ray emission from the burst can be studied simultaneously with the
X-ray and UV/optical emission.  The data from the BAT will also produce
a sensitive hard X-ray all-sky survey over the course of Swift's two
year mission (see~\ref{section:science:hardsurvey}).  Figure~\ref{fig4}
shows a cut-away drawing of the BAT, and Table~\ref{tab2} lists the
BAT's parameters.  Further information on the BAT is given by
\citet{barthelmy2003}.

\subsection{Technical Description} \label{section:bat:technical}

The BAT's 32,768 pieces of $4 \times 4 \times 2$ mm CdZnTe (CZT) form a
$1.2 \times 0.6$ m sensitive area in the detector plane.  The detector
configuration is similar to that of the CdTe detectors on INTEGRAL, with
about twice the area of the INTEGRAL array.  Groups of 128 detector
elements are assembled into $8 \times 16$ arrays, each connected to
128-channel readout Application Specific Integrated Circuits (ASICs; the
XA1s, which are designed and produced by Integrated Detector and
Electronics of Norway). Detector modules, each containing two such
arrays, are further grouped by eights into blocks.  This hierarchical
structure, along with the forgiving nature of the coded-aperture
technique, means that the BAT can tolerate the loss of individual
pixels, individual detector modules, and even whole blocks without
losing the ability to detect bursts and determine locations. The CZT
array will have a nominal operating temperature of 20\degr C, and its
thermal gradients (temporal and spatial) will be kept to within $\pm
1$\degr C.  The typical bias voltage is $Ü200$ V, with a maximum of
$Ü300$ V.  The detectors will be calibrated in flight with an electronic
pulsar and an $^{241}$Am tagged source.

The BAT has a D-shaped coded mask, made of $\sim 54,000$ lead tiles ($5
\times 5 \times 1$ mm) mounted on a 5 cm thick composite honeycomb
panel, which is mounted by composite fiber struts 1 meter above the
detector plane.  Because the large FOV requires the aperture to be much
larger than the detector plane and the detector plane is not uniform due
to gaps between the detector modules, the BAT coded mask uses a
completely random, 50\% open-50\% closed pattern, rather than the
commonly used Uniformly Redundant Array pattern.  The mask has an area
of 2.7 m$^{2}$, yielding a half-coded FOV of 100\degr $\times$ 60\degr,
or 1.4 steradians.

A graded-Z fringe shield, located both under the detector plane and
surrounding the mask and detector plane, will reduce background from the
isotropic cosmic diffuse flux and the anisotropic Earth-albedo flux by
$\sim 95$\%.  The shield is composed of layers of Pb, Ta, Sn, and Cu,
which are thicker toward the bottom nearest the detector plane and
thinner near the mask.

An FoM algorithm resides within the BAT flight software and decides if a
burst detected by the BAT is worth requesting a slew maneuver by the
spacecraft.  If the new burst has more ``merit'' than the pre-programmed
observations, a slew request is sent to the spacecraft.  Uploaded
rapid-reaction positions are processed exactly the same as events
discovered by the BAT.  The FoM is implemented entirely in software and
can be changed either by adjusting the parameters of the current
criteria or by adding new criteria.

\subsection{BAT Operations} \label{section:bat:operations}

The BAT runs in two modes: Burst Mode, which produces burst positions,
and Survey Mode, which produces hard X-ray survey data.  In the Survey
Mode the instrument collects count-rate data in 5 minute time bins for
18 energy intervals.  When a burst occurs it switches into a
photon-by-photon mode with a round-robin buffer to save pre-burst
information.

\subsubsection{Burst Detection}

The burst trigger algorithm looks for excesses in the detector count
rate above expected background and constant sources.  It is based on
algorithms developed for the HETE-2 GRB observatory, upgraded based on
HETE-2 experience.  The algorithm continuously applies a large number of
criteria that specify the pre-burst background intervals, the order of
the extrapolation of the background rate, the duration of the burst
emission test interval, the region of the detector plane illuminated,
and the energy range.  The BAT processor will continuously track
hundreds of these criteria sets simultaneously.  The table of criteria
can be adjusted after launch.  The burst trigger threshold is
commandable, ranging from $4 \sigma$ to $11 \sigma$ above background
noise with a typical value of $8 \sigma$.  A key feature of the BAT
instrument for burst detection is its imaging capability.  Following the
burst trigger the on-board software will check for and require that the
trigger corresponds to a point source, thereby eliminating many sources
of background such as magnetospheric particle events and flickering in
bright galactic sources.  Time stamping of events within the BAT has a
relative accuracy of 100 $\mu$s and an absolute accuracy from the
spacecraft clock of $\sim 200$ $\mu$s (after ground analysis).  When a
burst is detected, the sky location and intensity will be immediately
sent to the ground and distributed to the community through the
Gamma-Ray Burst Coordinates Network (GCN) \citep{barthelmy2000}.

\subsubsection{Hard X-ray Survey}

While searching for bursts, the BAT will perform an all-sky hard X-ray
survey and monitor for hard X-ray transients.  The survey is described
in Section~\ref{section:science:hardsurvey}.  For on-board transient
detection, 1-minute and 5-minute detector plane count-rate maps and
$\sim 30$-minute long average maps are accumulated in 4 energy
bandpasses.  Sources found in these images are compared against an
on-board catalog of sources.  Those sources either not listed in the
catalog or showing large variability are deemed transients.  A subclass
of long smooth GRBs that are not detected by the burst trigger algorithm
may be detected with this process.  All hard X-ray transients will be
distributed to the world community through the internet, just like the
bursts.

\subsection{Detector Performance} \label{section:bat:performance}

\begin{figure}[ht]
   \figurenum{5}
   \epsscale{0.3}
   \plotone{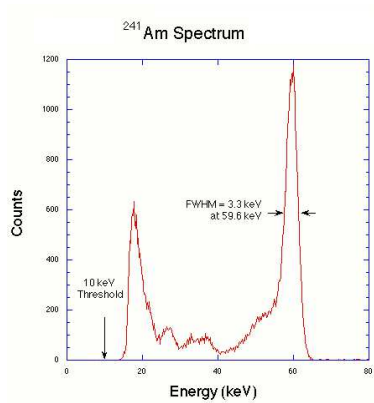}
   \caption{Typical spectrum of $^{241}$Am for a single CZT pixel.}
   \label{fig5}
\end{figure}

\begin{figure}[ht]
   \figurenum{6}
   \epsscale{0.5}
   \plotone{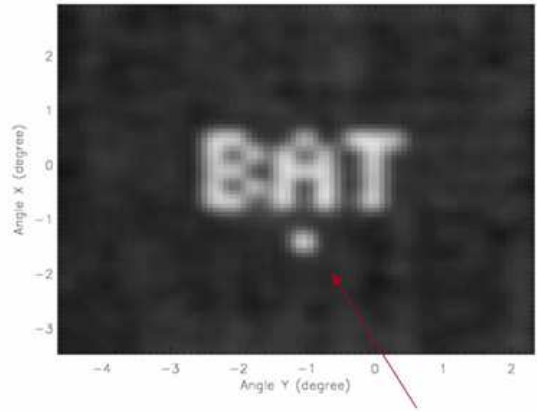}
   \caption{Composite image of a $^{133}$Ba radioactive gamma-ray
   emitter positioned at locations above the BAT instrument to spell
   ``BAT''.  The FWHM spread of the source image when corrected to
   infinite distance is 17 arcmin.}
   \label{fig6}
\end{figure}

A typical spectrum of the 60 keV gamma-ray line from an $^{241}$Am
radioactive source for an individual pixel is shown in
Figure~\ref{fig5}.  It has a full-width-half maximum (FWHM) at 60 keV of
3.3 keV ($\Delta E/E = 5\%$), which is typical of CZT detectors.  A
composite image made with a $^{133}$Ba radioactive calibration source is
shown in Figure~\ref{fig6}.  The source was positioned 3 meters above the
detectors and moved to several locations to spell out ``BAT''.  The FWHM
spread of the individual images, when corrected to infinite distance, is
17 arcmin, which is consistent with the predicted instrument PSF of
$<20$ arcmin. Simulations have calculated an average BAT background
event rate of 17,000 events s$^{-1}$, with orbital variations of a
factor of two around this value.  This yields a GRB sensitivity of $\sim
10^{-8}$ erg cm$^{-2}$ s$^{-1}$, 5 times better than BATSE.  The
combination of the 4 mm square CZT pieces, plus the 5 mm square mask
cells and the 1-m detector-to-mask separation gives an instrumental
angular resolution of 20 arcmin FWHM, yielding a conservative 4 arcmin
centroiding capability for bursts and steady-state sources given an $8
\sigma$ burst threshold.

\section{X-Ray Telescope}
\label{section:xrt}

\begin{figure}[ht]
   \figurenum{7}
   \epsscale{0.5}
   \plotone{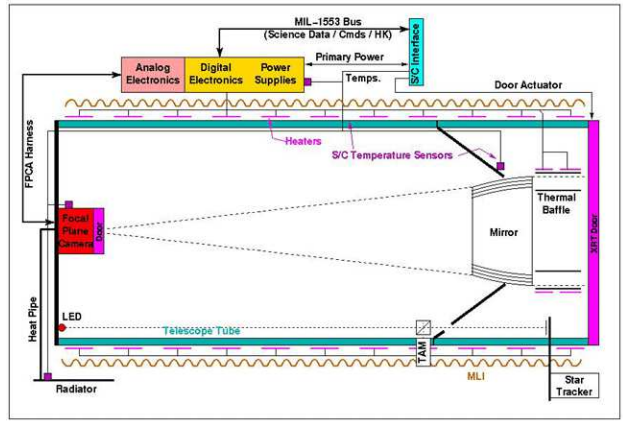}
   \caption{Block diagram of Swift's X-Ray Telescope}
   \label{fig7}
\end{figure}

Swift's X-Ray Telescope (XRT) is designed to measure the fluxes,
spectra, and lightcurves of GRBs and afterglows over a wide dynamic
range covering more than 7 orders of magnitude in flux.  The XRT will
pinpoint GRBs to 5-arcsec accuracy within 10 seconds of target
acquisition for a typical GRB and will study the X-ray counterparts of
GRBs beginning 20-70 seconds from burst discovery and continuing for
days to weeks.  Figure~\ref{fig7} shows a schematic of the XRT, and
Table~\ref{tab3} summarizes XRT parameters.  Further information on the
XRT is given by \citet{burrows2003}.

\subsection{Technical Description}
\label{section:xrt:technical}

The XRT is a focusing X-ray telescope with a 110 cm$^{2}$ effective
area, 23 arcmin FOV, 18 arcsec resolution (half-power diameter), and
0.2-10 keV energy range.

The XRT uses a grazing incidence Wolter 1 telescope to focus X-rays onto
a state-of-the-art CCD.  The complete mirror module for the XRT consists
of the X-ray mirrors, thermal baffle, a mirror collar, and an electron
deflector.  The X-ray mirrors are the FM3 units built, qualified and
calibrated as flight spares for the JET-X instrument on the Spectrum X
mission \citep{citterio1996, wells1992, wells1997}. To prevent on-orbit
degradation of the mirror module's performance, it will be maintained at
$20 \pm 5$\degr C with gradients of $< 1$\degr C by an actively
controlled thermal baffle similar to the one used for JET-X.

A composite telescope tube holds the focal plane camera, containing a
single CCD-22 detector.  The CCD-22 detector, designed for the EPIC MOS
instruments on the XMM-Newton mission, is a three-phase frame-transfer
device, using high resistivity silicon and an open-electrode structure
\citep{holland1996} to achieve a useful bandpass of 0.2-10 keV
\citep{short1998}. The CCD consists of an image area with $600 \times
602$ pixels ($40 \times 40$ mm) and a storage region of $600 \times 602$
pixels ($39 \times 12$ mm).  The FWHM energy resolution of the CCD
decreases from $\sim 190$ eV at 10 keV to $\sim 50$ eV at 0.1 keV, where
below $\sim 0.5$ keV the effects of charge trapping and loss to surface
states become significant.  A special ``open-gate'' electrode structure
gives the CCD-22 excellent low energy quantum efficiency (QE) while high
resistivity silicon provides a depletion depth of 30 to 35 mm to give
good QE at high energies.  The detectors will operate at approximately
$Ü100$ K to ensure low dark current and to reduce the CCD's
sensitivity to irradiation by protons (which can create electron traps
which ultimately affect the detector's spectroscopy).

\subsection{Operation and Control}
\label{section:xrt:operations}

The XRT supports three readout modes to enable it to cover the dynamic
range and rapid variability expected from GRB afterglows, and
autonomously determines which readout mode to use.  In order of bright
flux capability (and the order that would normally be used following a
GRB), the modes are as follows.  Imaging Mode produces an integrated
image measuring the total energy deposited per pixel and does not permit
spectroscopy.  It will only be used to position bright sources up to $7
\times 10^{-7}$ erg cm$^{-2}$ s$^{-1}$ (37 Crab).  Windowed Timing Mode
sacrifices position information to achieve high time resolution (2.2 ms)
and bright source spectroscopy through rapid CCD readouts.  It is most
useful for sources with flux below $\sim 10^{-7}$ erg cm$^{-2}$ s$^{-1}$
(5 Crab). Photon-counting Mode uses sub-array windows to permit full
spectral and spatial information to be obtained for source fluxes
ranging from $2 \times 10^{-14}$ to $2 \times 10^{-11}$ erg cm$^{-2}$
s$^{-1}$ (1 mCrab to 45 mCrab).  The upper limit is set by when  pulse
pile-up becomes important ($> 5\%$).

\subsection{Instrument Performance} \label{section:xrt:performance}

The instrument point spread function has a 18 arcsec half-energy width,
and, given sufficient photons, the centroid of a point source image can
be determined to sub-arcsec accuracy in detector coordinates.  Based on
BeppoSAX and RXTE observations, it is expected that a typical X-ray
afterglow will have a flux of 0.5-5 Crabs in the 0.2-10 keV band
immediately after the burst.  This flux should allow the XRT to obtain
source positions to better than 1 arcsec in detector coordinates, which
will increase to $\sim 5$ arcsec when projected back into the sky due to
alignment uncertainty between the star tracker and the XRT.

\begin{figure}[ht]
   \figurenum{8}
   \epsscale{0.3}
   \plotone{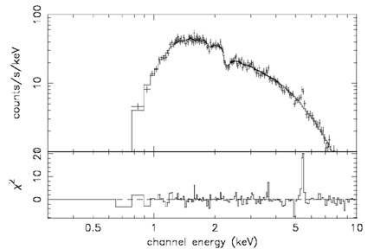}
   \caption{Simulated spectrum from 100 s XRT observation of a typical
   150 mCrab afterglow at z = 1.0, assuming a power law spectrum plus a
   Gaussian Fe line at 6.4 keV}
   \label{fig8}
\end{figure}

The XRT energy resolution at launch will be $\sim 140$ eV at 6 keV, with
spectra similar to that shown in Figure~\ref{fig8}.  Fe emission lines,
if detected, will provide a redshift measurement accurate to about 10\%.
The resolution will degrade during the mission, but will remain above
300 eV at the end of the mission life for a worst-case environment.
Photometric accuracy will be good to 10\% or better for source fluxes
from the XRT's sensitivity limit of $2 \times 10^{-14}$ erg cm$^{-2}$
s$^{-1}$ to $\sim 8 \times 10^{-7}$ erg cm$^{-2}$ s$^{-1}$ (about 2
times brighter than the brightest X-ray burst observed to date).

\section{Ultra-Violet/Optical Telescope}
\label{section:uvot}

\begin{figure}[ht]
   \figurenum{9}
   \epsscale{0.5}
   \plotone{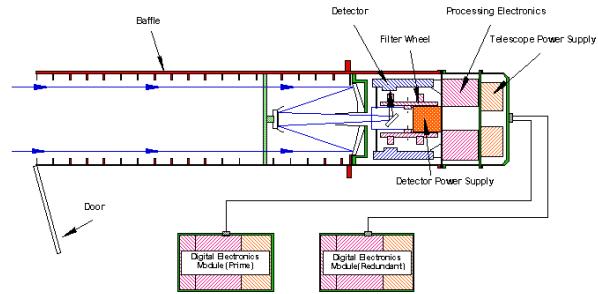}
   \caption{Diagram of Swift's Ultra-Violet/Optical Telescope.}
   \label{fig9}
\end{figure}

The Ultra-Violet/Optical Telescope (UVOT) design is based on the Optical
Monitor (OM) on-board ESA's XMM-Newton mission (see \citet{mason1996,
mason2001} for a discussion of the XMM-OM).  UVOT is co-aligned with the
XRT and carries an 11-position filter wheel, which allows low-resolution
grism spectra of bright GRBs, and broadband UV/visible photometry.
There is also a $4\times$ field expander (magnifier) that delivers
diffraction limited sampling of the central portion of the telescope
FOV.  Photons register on the microchannel plate intensified CCD (MIC).
Figure~\ref{fig9} shows a schematic of the UVOT, and Table~\ref{tab4}
summarizes the UVOT parameters.  Further information on the UVOT is
given by \citet{roming2003}.

\subsection{Technical Description}
\label{section:uvot:technical}

The UVOT's optical train consists of a 30 cm clear aperture
Ritchey-ChrÚtien telescope with a primary f-ratio of f/2.0 increasing to
f/12.72 after the secondary.  The baffle system consists of an external
baffle, which extends beyond the secondary mirror; an internal baffle,
which lines the telescope tube between the primary and secondary
mirrors; and primary/secondary baffles, which surround the secondary
mirror and the hole at the center of the secondary mirror.  An INVAR
structure that is intrinsically thermally stable is used between the
mirrors and maintains the focus.  Fine adjustment to the focus is
achieved by activating heaters on the secondary mirror support structure
and on the INVAR metering rods that separate the primary and secondary
mirrors.

The UVOT carries two redundant photon-counting detectors that are
selected by a steerable mirror mechanism.  Each detector has a filter
wheel mounted in front of it carrying the following elements: a blocked
position for detector safety; a white light filter; a field expander;
two grisms; U, B, and V filters; and three broadband UV filters centered
on 190, 220 and 260 nm.  One grism on each wheel is optimized for the
UV, the other for optical light, and both offer a spectral resolution of
$\sim 1$ nm/pixel.  Diffraction-limited images can be obtained with the
$4\times$ field expander (magnifier); however, because of the limits of
the transmission optics, the magnifier does not work at UV wavelengths.
The UVOT operates as a photon-counting instrument.  The two detectors
are MICs incorporating CCDs with $384 \times 288$ pixels, $256 \times
256$ of which are usable for science observations.  Each pixel
corresponds to $4 \times 4$ arcsec on the sky, providing a $17 \times
17$ arcmin FOV. Photon detection is performed by reading out the CCD at
a high frame rate and determining the photon splash's position using a
centroiding algorithm.  The detector achieves a large format through
this centroiding technique, sub sampling the $256 \times 256$ CCD pixels
each into $8 \times 8$ virtual pixels, leading to an array of $2048
\times 2048$ virtual pixels with a size of $0.5 \times 0.5$ arcsec on
the sky.  The frame rate for the UVOT detectors is 10.8 ms. These
detectors have very low dark current, which usually can be ignored when
compared to other background sources.  In addition, they have few hot or
dead pixels and show little global variation in quantum efficiency.

\subsection{Operating Scenarios}
\label{section:uvot:operations}

There are six observing scenarios for the UVOT: slewing, settling,
finding chart, automated targets, pre-planned targets, and safe pointing
targets.

{\em Slewing}.  As the spacecraft slews to a new target, the UVOT does
not observe in order to protect itself from bright sources slewing
across its FOV and damaging the detector.

{\em Settling}.  After notification from the spacecraft that the
intended object is within ten arcminutes of the target, the UVOT begins
observing.  During this phase pointing errors are off-nominal, i.e., the
target is moving rapidly across the FOV as the spacecraft settles.  The
positional accuracy is only known to a few arcmin based on the BATÍs
centroided position.

{\em Finding Chart}.  If the intended target is a new GRB and the
spacecraft is sufficiently settled, i.e., the pointing errors are small,
the UVOT begins a 100 second exposure in the V filter to produce a
finding chart.  The finding chart is to aid ground-based observers in
localizing the GRB.  The positional accuracy of the finding chart will
be approximately 0.3 arcsec relative to the background stars in the FOV.
It is anticipated that for most bursts the XRT will have reported a
better than 5-arcsec position for the target before the end of the
finding chart observation.

{\em Automated Targets}.  Once a finding chart has been produced, an
automated sequence of exposures, which uses a combination of filters, is
executed.  The sequence is based on the optical decay profile of the GRB
afterglow and time since the initial burst.  Currently, two automated
sequences will be launched:  bright and dim GRB sequences.  Although
only two sequences will be loaded at launch, new sequences can be added
and existing ones modified as GRB afterglows become better understood.

{\em Pre-Planned Targets}.  When there are no automated targets,
observation of planned targets (which have been uploaded to the
spacecraft) begins.  Follow-up of previous automated targets,
targets-of-opportunity, and survey targets are included as planned
targets.

{\em Safe Pointing Targets}.  When observing constraints do not allow
observations of automated or pre-planned targets the spacecraft points
to predetermined locations on the sky that are observationally safe for
the UVOT.

There are two data collection  modes for the UVOT: Event and Imaging,
which can be run at the same time if desired. In Event Mode, the UVOT
stores time-tagged photon events in memory as they arrive. The timing
resolution is equal to the CCD frame time ($\sim 11$ ms). In Imaging
Mode, photon events are summed into an image for a time period known as
the tracking frame time ($\leq 20$ s).  These tracking frame images are
shifted to compensate for spacecraft pointing drift and added in memory,
using the bright stars within the FOV as fiducial points.  The advantage
of Imaging Mode is that it minimizes the telemetry requirements when the
photon rate is high, but at the expense of timing information. The area
of the sky over which data are stored can be windowed in each mode,
again allowing the optimum utilization of telemetry.  In general it is
anticipated that a large window will be used during the initial phases
of the burst when the uncertainty in its position on the detector is
higher, and that the window size will be reduced when the burst is
positioned more accurately.

Besides the science data collection modes, the UVOT also supports a
number of engineering modes to monitor on-orbit performance and aid
instrument testing and integration.

\subsection{Instrument Performance}
\label{section:uvot:performance}

The top-level UVOT observational capabilities are to provide information
on the short-term UV/optical behavior of GRBs, a finding chart for
ground-based observers, and GRB follow-up observations.  The UVOT must
also be able to protect itself autonomously from bright sources that
could damage the detectors.

The 100 second parameterized finding chart is used to construct an image
on the ground and provide a $\sim 0.3$ arcsec position for the burst
relative to the field stars close to the GRB.  Redshifts can be obtained
for brighter GRBs with grism spectra. For fainter GRBs, light curves
positions, and photometric redshifts will be obtained by cycling through
the 6 broadband filters.  Event Mode data will allow monitoring of
source variability on short timescales. Centroided positions accurate to
0.3 arcsec will be determined, allowing the UVOT to accurately position
the burst relative to any host galaxy it may be associated with.  The
UVOT will have a $5 \sigma$ sensitivity to a limiting magnitude of
$B = 24.0$ in 1000 s using the white light filter.

The UVOT provides for its own detector safety to a greater extent than
OM, as it must autonomously and quickly respond to new burst detections.
If the detector is in danger of being overexposed, this can be rapidly
sensed by circuitry in the camera head, which signals the instrument
control unit (ICU) to drop the voltage on the detector photocathode,
rendering it insensitive.  A catalog of bright sources will be included
in the ICU.  This will be consulted whenever a slew is triggered to
anticipate the presence of known objects in the new FOV that might
damage the detector  and limit the exposure to a safe value.

\section{Ground System}
\label{section:ground}

\begin{figure}[ht]
   \figurenum{10}
   \epsscale{0.75}
   \plotone{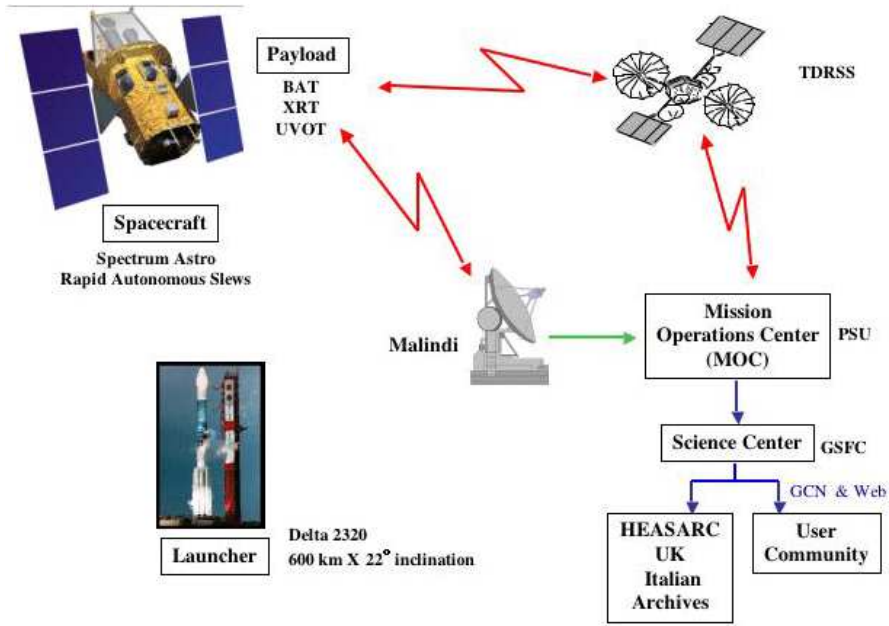}
   \caption{The Swift mission architecture.}
   \label{fig10}
\end{figure}

Swift's ground system has been designed for speed and flexibility both
in distributing burst alerts and data and in receiving scientific input
for mission planning.  The Swift ground system consists of Penn State
University's (PSU's) Mission Operation Center (MOC), the Swift Science
Center (SSC) and Swift Data Center (SDC) at GSFC, the UK Swift Science
Data Center (UKSSDC) at the University of Leicester, the Italian Swift
Archive Center (ISAC), the Italian Space Agency's (ASI) ground station at
Malindi, Kenya, the NASA TDRSS data relay satellites, and communications
networks interfacing the various elements.  The overall mission
architecture is shown in Figure~\ref{fig10}.

Swift burst alerts and burst characteristics are transmitted almost
instantaneously through a TDRSS link to the GCN for rapid distribution
to the astronomical community.  A TDRSS uplink also permits rapid
response to ToOs, such as GRBs detected by other missions.

The Malindi ground station in Kenya provides the primary communications
support.  There, telemetry frames are timestamped and sorted by channel.
  Real-time data are forwarded immediately to the MOC.  The remainder of
the telemetry is stored temporarily and later sent to the MOC.  Malindi
is also the primary station for forwarding commands from the MOC to the
spacecraft.

The MOC, located near the PSU campus in State College PA, provides
real-time command and control of the spacecraft and monitors the
observatory.  In addition, the MOC takes care of science and mission
planning, ToO handling and data capture and accounting.

The SDC is part of the Astrophysics Data Facility in the Space Science
Data Operations Office at GSFC.  It makes Level 0, 1, 2, and 3 data
products, and then provides the data to the High Energy Astrophysics
Science Archive Research Center (HEASARC), ISAC and UKSSDC.  The
HEASARC, which is part of the Laboratory for High Energy Astrophysics
(LHEA) at GSFC, is responsible for making the data available to the
astronomical community and for the long-term archive of the data.  The
data centers in the UK and Italy distribute data to their respective
communities.

The SSC supports the science community in the use of Swift data with
documentation and advice.  The SSC is also responsible for developing
the data analysis tools for the UVOT and takes the lead role in
integrating the entire suite of science analysis tools for Swift data.
Data analysis tools for the XRT and BAT data will be developed by the
ISAC and BAT instrument teams respectively.

\section{Follow-Up Team}
\label{section:followup}

An essential aspect of the Swift mission is the ability to make hundreds
of burst positions, as well as the positions of transient sources
detected during the sky survey, available to the wide community for
ground- and space-based multiwavelength follow-up studies.  Up to now,
such studies have been conducted largely on a ToO basis, since only
$\sim 20$ bursts per year have had their positions determined well
enough to observe with large telescopes.  This will change when Swift
flies, because the number of positions will be sufficient to propose a
routine counterpart observing program, with the assurance that whenever
observing time is granted, there will be an interesting, recent event to
observe.

We will encourage both ToO and routine observations by distributing
precise positions in near-real-time via the GCN and any other means that
are in use during the mission.  As is the case today, observers will
have completely free access to the public data.  In order to maximize
the routine use of the most sensitive multiwavelength and
non-electromagnetic detectors over the widest possible geographic range,
we have formed a team of over 30 scientists (see Table~\ref{tab5}) who
will collaborate closely with the project on follow-up observations.
They have, or will request, access to instruments, including the worldÍs
largest telescopes, to carry out high-resolution spectroscopy, optical,
IR and radio monitoring of light curves, and morphological studies of
host galaxies, for example.  We will encourage all observers to make
their data public by posting it to a web-accessible database, which we
will maintain, and which will centralize all known observations of each
burst.

\section{Guest Investigator Program}
\label{section:gi}

It is anticipated that there will be substantial interest and
involvement of the astronomical community in the Swift mission,
particularly in the areas of GRB and other transient-source science.
The data from the mission will be made public as soon as they are
processed to allow as many researchers as possible to participate.
Also, the near-real-time distribution of alerts for GRBs and other
transients through the GCN will facilitate prompt follow-up observations
by ground and space instruments.

Support for community involvement will be provided by a NASA Guest
Investigator (GI) program.  Proposals will be solicited through the
Research Opportunities for Space Science (ROSS) yearly solicitation.
The program will be open to investigations with US principal
investigators.  Areas of research solicited in the first year of the
program will be the following:

\begin{itemize}
	\item Correlative observations of GRBs with non-Swift instruments
	and observatories

	\item New GRB projects not duplicative of Swift team key projects
	and not requiring GI-specified observatory pointings

	\item Theoretical investigations that advance the mission science
	return in the area of GRBs
\end{itemize}

The program may be expanded in year 2 and later to include areas of
research in non-GRB science and may include GI-specified pointings,
depending on the experience gained from year 1 of the mission.

For year 1, GI proposals were due in fall 2003 with investigations
starting 4.5 months after launch.  There will be approximately 30
investigations selected.

\section{Swift Data Processing and Multiwavelength Analysis}
\label{section:data}

When Swift telemetry is received from the MOC at the SDC, it triggers a
run of the Swift processing pipeline -- a detailed script of tasks that
produce Flexible Image Transportation System (FITS) files from raw telemetry, calibrated event lists and
cleaned images, and higher level science products such as light curves
and spectra for all Swift instruments. Initial data products appear on
the Swift Quick Look Data public Web site. When processing is complete,
the products are delivered to the HEASARC archive. All pipeline software
will be FITS Tools (FTOOLs), that will also be distributed to Swift users. This way,
users can then reprocess/reanalyze data when new calibration information
is made available, instead of needing to wait for eventual reprocessing.

Software tools specific to the BAT, XRT and UVOT will apply
instrument-specific calibration information and filtering criteria in
the pipeline to arrive at calibrated images and screened event lists.

\subsection{UVOT pipeline}
\label{section:data:uvot}

The UVOT instrument produces a finding chart which arrives via TDRSS,
and event and image data taken through any one of six broadband filters
or two grism filters. During an observation, the size and location of
the window can change, as can the on-chip binning. The pipeline will
produce cleaned, calibrated event list files, calibrated image files and
standard products for each observation.  This includes, e.g., high
signal-to-noise images of the field generated by combining all
individual images obtained using the same filter. Exposure maps are
constructed for each combined image. Source lists are derived from the
combined images. Provided an optical counterpart to the target has been
identified, light curves for each available filter are extracted from
all available image and event data.

Grism spectra of the candidate counterpart are obtained from each
available grism image. Grism source event tables will be generated,
containing the wavelength of each photon, screened according to a
spatial mask so that only those events likely to be associated with the
candidate counterpart remain. A response matrix will be generated
to facilitate the analysis of grism data within the XSPEC software package. A "response
matrix" will also be provided so that broad-band fluxes through the
standard filters can be fit simultaneously with XRT and BAT data.

\subsection{XRT Pipeline}
\label{section:data:xrt}

The XRT instrument has several readout modes to cover the large dynamic
range and rapid variability expected from GRB afterglows, and is capable
of autonomously changing the readout mode when the source flux changes
appropriately. For all XRT data modes, the XRT pipeline will produce
cleaned, calibrated event list files and standard products for each
observation.

The standard event screening criteria for XRT will rely on event grade,
detector temperature, and spacecraft attitude information.  XRT
standard products -- spectra, images and light curves -- will be produced
in the pipeline, as well as Ancillary Response File (ARF) files for spectroscopy, an exposure map
appropriate for converting counts in an image to flux.

Computation of an appropriate exposure map for images will result in the
net exposure time per pixel taking into account attitude reconstruction,
spatial quantum efficiency, filter transmission, vignetting and FOV. 
For spectra, a Redistribution Matrix File (RMF) that specifies the
channel probability distribution for a photon of a given energy and an
ARF specifying telescope effective area and
window absorption will be calculated in the pipeline.

\subsection{BAT Pipeline}
\label{section:data:bat}

The BAT instrument produces event files for each BAT event data type
(long, short, long calibration, short calibration), as well as rate
files for the entire BAT array, mask-tagged light curves for each of
three mask-tagged sources, and light curves of GRBs. The BAT also
produces Detector Plane Histograms (DPHs) in survey mode. The pipeline produces
standard Burst Mode data products: burst spectra on various time scales,
response matrices, light curves, and images. Raw BAT DPHs are used to generate the BAT survey products. Some
survey products are produced for a single DPH, and others result from
summing all DPHs in a given pointing. Light curves are generated for
survey sources applying a source detection algorithm, the mask pattern
and a cleaning algorithm to eliminate confusing sources.

For the entire BAT array, a one-second light curve will be produced,
while light curves with 64 ms time resolution are produced for entire
BAT blocks.  Light curves with 1.6-s time resolution in four energy
ranges are produced for each array quadrant.  Light curves with 8-s
time resolution in four energy bands record the maximum count rate in
each of nine array regions, on 5 time scales.  Mask-tagged light curves
with 1.6-s time resolution are generated for each of three mask-tagged
sources.  Light curves of GRBs derived from event data and five-minute
light curves derived from the survey data for each source detected by
the BAT are also generated in the pipeline.

The pipeline produces BAT event files and Detector Plane Images (DPIs),
needed to generate sky images. DPIs are histogram images of calibrated
events, and must be deconvolved with the mask to produce useable sky
images.  Event files are rebinned to produce burst light curves and
spectra. Photon spectra may be derived by fitting count spectra and can
be corrected for the effects of partial coding and the reduced off-axis
response. Detector response matrices are also calculated in the
pipeline.

For BAT survey data, count spectra and response matrices will be
produced and archived for sources found in the survey.  During burst
mode, count spectra and response matrices will be generated for bursts
before, during and after the slew.  Photon spectra before, during and
after the slew may also be produced during burst mode. The pipeline will
produce deconvolved sky images containing data from entire snapshots.
Images will be available in four energy bands as well as a broadband image.

\subsection{Multi-wavelength Analysis}
\label{section:data:multiwavelength}

SwiftÕs unique purpose of observing a GRB in progress and its afterglow
across the spectrum demands that spectral data from the three
instruments be analyzed together.  With this goal in mind, UVOT spectra
are prepared with the associated response matrices and calibration
information to allow them to be input into XSPEC (as well as the more
traditional UV-optical spectral analysis tools, such as IRAF).  Assuming
a well-understood cross calibration of the instruments, such
functionality will allow users to fit various spectral models for
bursts, search for the energy range of spectral breaks, and accurately
measure the Ly-$\alpha$\  cut-off (and hence the redshift) for the
afterglow.

\section{Mission Implementation}
\label{section:implementation}

Swift is being developed by an international collaboration with primary
hardware responsibilities at GSFC, Penn State, Mullard Space Science
Laboratory, University of Leicester, Osservatorio Astronomico di Brera
and the ASI Science Data Center.  Other institutions that have made
significant contributions to the mission are Los Alamos National
Laboratory (LANL), Sonoma State University, Max Planck Institut fÄr
Extraterrestriche Physik, and Institute of Space and Astronautical
Science in Japan.

The mission is managed at GSFC.  Swift's BAT instrument is being
developed at GSFC with flight software from LANL.  The XRT is a
collaboration between Penn State University, University of Leicester and
Osservatorio Astronomico di Brera.  Groups at Mullard Space Science
Laboratory and Penn State University have developed the UVOT.  The
Follow-up Team has been formed under leadership at UC Berkeley to
perform follow-up observations of Swift-detected GRBs at other
observatories.  An Education and Public Outreach (EPO) team is in place
under the direction of Sonoma State University.  The responsibilities of
the various institutions involved in the Swift mission are listed in
Table~\ref{tab6}.

The mission team successfully completed the Critical Design Review (CDR)
in July 2001 and the Mission Operations Review (MOR) in August 2002.
Instrument deliveries will complete by February 2004.  Launch is
scheduled for mid 2004.

Following launch, the first 45 days in orbit will be the observatory
activation and checkout phase during which time the instruments and
spacecraft are turned on and tested.  The spacecraft slew testing will
occur starting at approximately 14 days and the doors of the XRT and
UVOT will be opened at approximately 25 days.  After the activation and
checkout phase, the next 3 months will be the verification phase during
which time the observatory performance and data products will be
verified and the instruments calibrated.  The Swift team anticipates
that BAT GRB alerts will be distributed to the community over the GCN
starting after the activation and checkout phase.  The initial alerts
will be sent out several hours after the GRB onset to allow ground
verification of the data.  By the end of the verification phase, the
observatory will be up to full performance with GRB data from all
instruments rapidly distributed via the GCN.

\section{Education and Public Outreach}
\label{section:epo}

The science from the Swift mission appeals to students of all ages, who
are naturally excited about GRBs and black holes.  The Swift EPO program
capitalizes on that existing interest to teach basic physical science
(e.g., the electro-magnetic spectrum, gravity, and the cycles of energy
and matter) as well as more advanced Swift science (e.g., GRBs, black
holes, and cosmology.)

The Swift EPO program includes partners from the US, Italy, Germany and
the UK. It is divided into 5 basic program elements: Swift web site,
printed materials, educator training, informal education and program
evaluation.   In addition, a unique element of the Swift outreach
program is that a song has been written and recorded about the science
and mission that can be heard at the Swift main web site
(\url{http://swift.gsfc.nasa.gov}) or the EPO web site give below.

{\em Swift EPO Web Site}.  The EPO website (\url{http://swift.sonoma.edu})
contains an overview of the Swift mission and the EPO program, as well
as downloadable versions of all Swift educational materials.  It is
updated frequently, and is linked to the sites of the international EPO
partners.

{\em Printed Materials}.  The Swift EPO effort has already produced several
printed products, including posters outlining NewtonÍs three laws,
booklets with classroom activities based on the electromagnetic spectrum
and waves, and a deck of cards designed to teach students scientific
notation.   A major product is the Great Explorations in Math and
Science (GEMS) guide titled ñThe Invisible Universe: the Electromagnetic
Spectrum from Radio Waves to Gamma-rays.î  This well-tested classroom
workbook contains a series of five hands-on activities that use the
mystery of gamma-ray bursts to teach the electromagnetic spectrum.  It
was developed in partnership with the GEMS group at the Lawrence Hall of
Science.

{\em Educator Training}.  The Swift EPO team has held several teacher
workshops at various national and local venues, training teachers to use
the Swift educational materials.  Workshops will be held throughout the
duration of the Swift mission, using new materials as they are
developed.  Swift materials are featured in a yearly summer school held
at the Pennsylvania State University, which includes many teachers from
rural, under-represented schools.  Recently, two Swift Education
Ambassadors (EAs) were appointed to help develop, evaluate and
disseminate Swift educational materials.  These award-winning educators
were selected in a national search, joining an ever-growing contingent
of EAs that are supporting other NASA missions.

{\em Informal Education}.  Swift materials are being incorporated as part of
the ñCosmic Questionsî museum exhibit being developed by the NASAÍs
Structure and Evolution of the Universe Education Forum and in an
exhibit at the British National Space Science Center.  Swift also
sponsors news briefs on the television show ñWhatÍs in the News?î  This
show reaches millions of middle school students each year, airing
through the Penn State public television network, WPSX. Three or four
segments each year feature the science and technology of the Swift
mission, as well as interviews with Swift scientists.

{\em Program Evaluation}.  The Swift educational materials are being
thoroughly assessed to ensure they are interesting, effective, widely
disseminated and aligned with the National Science and Mathematics
Education Standards.  The Swift Education Committee Ü a team comprised
of Swift scientists and master educators Ü and the EAs formatively
evaluate the materials as they are developed, while WestEd performs the
summative external program assessment, surveying workshop participants
and providing written analyses of the effectiveness of all materials and
workshops.

\section{Conclusion}

Swift is an innovative mission for gamma-ray burst study that will carry
three instruments to perform multiwavelength, simultaneous observations
of GRBs in gamma-ray, X-ray, and UV/optical wavelengths.  The Swift
spacecraft will be capable of rapid, autonomous slewing to capture
afterglows in the first minutes following a GRB.  New technology will
allow an advanced gamma-ray detector to image GRBs at 5 times better
sensitivity than BATSE.  Data will be rapidly distributed throughout the
scientific community, and participation by observers around the world is
encouraged.

\acknowledgments
We are greatly indebted to the management, engineering and support teams
who have worked tirelessly for the past 3 years to bring the Swift
mission to fruition and to NASA, PPARC and ASI for funding of the Swift
program in the US, the UK and Italy, respectively.

\pagebreak

\begin{deluxetable}{ll}
\tablecolumns{2}
\tabletypesize{\normalsize}
\tablecaption{Swift Mission Characteristics}
\tablewidth{0pt}
\tablehead{
\colhead{Mission Parameter} &
\colhead{Value}
}
\startdata
Slew Rate			& 50\degr in $< 
75$ s 
	\\
Orbit				& Low Earth, 600 
km altitude 				\\
Inclination			& 22\degr 
 
				\\
Launch Vehicle	& Delta 7320-10 with 3 meter fairing 	\\
Mass					& 1450 kg 
 
					\\
Power				& 1040 W 
 
				\\
Launch Date		& early 2004
\enddata
\label{tab1}
\end{deluxetable}

\begin{deluxetable}{ll}
\tablecolumns{2}
\tabletypesize{\normalsize}
\tablecaption{Burst Alert Telescope Characteristics}
\tablewidth{0pt}
\tablehead{
\colhead{BAT Parameter} &
\colhead{Value}
}
\startdata
Energy Range 
				& 15-150 keV 
 
				\\
Energy Resolution 
			&$\sim 7$ keV 
 
		\\
Aperture 
					& Coded 
mask, random pattern, 50\% open	\\
Detection Area 
			& 5240 cm$^{2}$ 
 
		\\
Detector Material 
			& CdZnTe (CZT) 
 
		\\
Detector Operation 
			& Photon counting 
 
		\\
Field of View (FOV) 
			& 1.4 sr (half-coded) 
 
	\\
Detector Elements 
			& 256 modules of 128 
elements/module		\\
Detector Element Size 
		& $4 \times 4 \times 2$ mm$^{3}$ 
			\\
Coded-Mask Cell Size 
		& $5 \times 5 \times 1$ mm$^{3}$ 
Pb tiles	\\
Telescope PSF 
				& $< 20 arcmin$ 
 
				\\
Source Position and Determination		& 
1-4 arcmin 
						\\
Sensitivity 
				& $\sim 10^{-8}$ 
erg cm$^{-2}$ s$^{-1}$	\\
Number of Bursts Detected				& $> 100$ yr$^{-1}$
\enddata
\label{tab2}
\end{deluxetable}

\begin{deluxetable}{ll}
\tablecolumns{2}
\tabletypesize{\normalsize}
\tablecaption{X-Ray Telescope Characteristics}
\tablewidth{0pt}
\tablehead{
\colhead{XRT Parameter} &
\colhead{Value}
}
\startdata
Energy Range				& 0.2-10 
keV

			\\
Telescope					& 
JET-X Wolter 1

		\\
Detector					& 
E2V CCD-22

			\\
Effective Area			& 110 cm$^{2}$ @ 
1.5 keV 
 
					\\
Detector Operation		& Photon 
counting, integrated imaging, and timing 
 
		\\
Field of View (FOV)		& $23.6 \times 
23.6$ arcmin 
 
				\\
Detection Elements		& $600 \times 
602$ pixels 
 
					\\
Pixel Scale					& 
2.36 arcsec

				\\
Telescope PSF				& 18 
arcsec HPD @ 1.5 keV 
 
						\\
Sensitivity					& 
$2 \times 10^{-14}$ erg cm$^{-2}$ s$^{-1}$ (1 
mCrab) in 10$^{4}$ s
\enddata
\label{tab3}
\end{deluxetable}

\begin{deluxetable}{ll}
\tablecolumns{2}
\tabletypesize{\normalsize}
\tablecaption{UltraViolet/Optical Telescope Characteristics}
\tablewidth{0pt}
\tablehead{
\colhead{UVOT Parameter} &
\colhead{Value}
}
\startdata
Wavelength Range		& 170-600 nm 
 
	\\
Telescope					& 
Modified Ritchey-Chr\'{e}tien	\\
Aperture					& 
30 cm diameter 
	\\
F-number					& 
12.7 
				\\
Detector					& 
Intensified CCD 
	\\
Detector Operation		& Photon counting 
						\\
Field of View (FOV)		& $17 \times 17$ 
arcmin				\\
Detection Elements		& $2048 \times 2048$ pixels		\\
Telescope PSF				& 0.9 
arcsec FWHM @ 350 nm		\\
Colors 						& 
6 
					\\
Sensitivity					& 
B = 24 in white light in 1000 s	\\
Pixel Scale					& 0.5 arcsec
\enddata
\label{tab4}
\end{deluxetable}

\begin{deluxetable}{lll}
\tablecolumns{3}
\tabletypesize{\small}
\tablecaption{Follow-Up Team}
\tablewidth{0pt}
\tablehead{
\colhead{Member} &
\colhead{Institution} &
\colhead{Facility of Expertise}
}
\startdata
Antonelli, Angelo			& 
Osservatorio Astronomico di Roma 
		& VLT, REM, FAME 
	\\
Bo\"{e}r, Michel			& CESR 
Toulouse 
						& 
TAROT 						\\
Buckley, David			& South Africa 
Astronomical Obs. 
	& SALT 
		\\
Busby, Michael			& Tennessee State 
University 
	& TSU telescopes 			\\
Canterna, Ron				& U. 
Wyoming 
						& 
WIRO 						\\
Cimatti, Andrea			& Osservatorio 
Arcetri, Florence 
	& LBT 
		\\
Coe, Malcolm				& U. 
Southampton 
					& 
Tenerife, IRTF, SAAO	\\
Courvoisier, Thierry	& INTEGRAL Science Data 
Centre				& INTEGRAL 
				\\
Covino, Stefano			& Osservatorio 
Astronomico di Brera			& VLT, 
REM 					\\
Della Valle, Massimo	& Osservatorio Arcetri, 
Florence					& 
La Silla, Paranal 			\\
Dingus, Brenda			& Los Alamos 
National Laboratory 
	& Milagro, VERITAS		\\
Fillippenko, Alex			& UC Berkeley 
 
			& KAIT, Keck 
			\\
Finn, Sam	 				& 
Penn State University 
				& LIGO 
				\\
Fiore, Fabrizio			& Osservatorio 
Astronomico di Roma				& 
ESO 
	\\
Fruchter, Andy			& STScI 
 
					& HST 
						\\
Ghisellini, Gabriele		& Osservatorio 
Astronomico di Brera			& VLT 
						\\
Gilmozzi, Roberto		& European 
Southern Observatory 
	& VLT 
		\\
Kawai, Nobuyuki			& RIKEN 
 
					& Okayama 
Observatory	\\
Kulkarni, Shri				& Caltech

	& Keck 
		\\
Margon, Bruce				& STScI 
 
						& 
HST 
	\\
Mundell, Carol				& John 
Moores U. 
					& 
Liverpool Telescope 		\\
Park, Hye-Sook			& Lawrence 
Livermore National Lab.				& 
Super-LOTIS 				\\
Pederson, Holger			& 
Copenhagen University Observatory 
	& NOT La Palma 				\\
Rhoads, James				& STScI 
 
						& 
KPNO, CTIO, IRTF 		\\
Schaefer, Brad			& U. Texas Austin 
 
				& McDonald, WIYN 
		\\
Schneider, Don			& Penn State 
University 
			& HET 
				\\
Skinner, Mark				& Boeing

	& AEOS 
	\\
Smith, Ian					& 
Rice University 
					& IR \& 
sub-mm, AEOS 	\\
Stubbs, Chris				& U. 
Washington 
						& 
ARC 
	\\
Thompson, Chris			& CITA 
 
						& 
SOAR 						\\
Vrba, Fred					& 
US Naval Observatory 
				& USNO telescopes 
		\\
Walton, Nic				& 
Institute of Astronomy, Cambridge 
		& INT 
			\\
Wheatley, Peter			& U. Leicester 
 
				& WASP, Faulkes 
			\\
Zerbi, Filippo				& 
Osservatorio Astronomico di Brera 
	& REM 
		\\
\enddata
\label{tab5}
\end{deluxetable}

\begin{deluxetable}{ll}
\tablecolumns{2}
\tabletypesize{\small}
\tablecaption{Swift Mission Responsibilities}
\tablewidth{0pt}
\tablehead{
\colhead{Responsibilities} &
\colhead{Lead Institution\tablenotemark{\dag}}
}
\startdata
Principal Investigator, Mission Management 
					& GSFC 
						\\
Spacecraft

	& Spectrum Astro			\\
BAT Instrument 
 
					& 
 
		\\
\hspace{0.1in}Management, Hardware 
						& 
GSFC 
	\\
\hspace{0.1in}On-board GRB Software 
						& 
LANL 
	\\
XTR Instrument 
 
					& 
 
		\\
\hspace{0.1in}Management, Electronics, Science 
Software	& PSU 
			\\
\hspace{0.1in}Detector System 
 
	& UL 
			\\
\hspace{0.1in}Mirrors 
 
				& OAB 
					\\
\hspace{0.1in}Calibration 
 
				& MPE 
					\\
UVOT Instrument 
 
				& 
 
	\\
\hspace{0.1in}Management, Electronics 
						& 
PSU 
	\\
\hspace{0.1in}Instrument Development 
						& 
MSSL 
	\\
Mission Integration and Test 
 
		& Spectrum Astro/GSFC	\\
Ground System Management 
 
			& GSFC 
				\\
Ground Station 
 
					& ASI 
						\\
Mission Operations Center 
 
			& PSU 
				\\
Science Center 
 
					& GSFC 
						\\
Data Centers 
 
						& 
GSFC, ASI/OAB, LU	\\
GRB Follow-up 
 
						& 
UCB 
	\\
Education/Public Outreach 
 
			& SSU
\enddata
\label{tab6}

\tablenotetext{\dag}{Abbreviations used in table:
\hspace{0.1in}ASI = Italian Space Agency;
\hspace{0.1in}GSFC = Goddard Space Flight Center;
\hspace{0.1in}LANL = Los Alamos National Laboratory;
\hspace{0.1in}MPE = Max Planck Institude fÄr Extraterrestrische Physik;
\hspace{0.1in}MSSL = Mullard Space Science Laboratory;
\hspace{0.1in}OAB = Osservatorio Astronomico di Brera;
\hspace{0.1in}PSU = Penn State University;
\hspace{0.1in}SSU = Sonoma State University;
\hspace{0.1in}UCB = University of California, Berkeley;
\hspace{0.1in}UL = University of Leicester}
\end{deluxetable}

\pagebreak

  \end{document}